\documentclass[letterpaper,final,superscriptaddress,tightenlines,raggedbottom,showpacs,nofootinbib,twoside,twocolumn,pdfstartview=FitH,10pt,
	amsmath,amssymb,amsfonts,aps,pra,floatfix,
	]{revtex4-1}

\allowdisplaybreaks

\usepackage[sort&compress]{natbib}
\usepackage[bookmarks,breaklinks,colorlinks,allcolors=blue,urlcolor=blue,linktocpage,hyperfootnotes,bookmarkstype=toc,bookmarksopen,bookmarksopenlevel=1,
	]{hyperref}
\hypersetup{pdfpagemode=UseNone}

\usepackage[T1]{fontenc}
\usepackage{wasysym}		
\usepackage{textcomp}
\usepackage{graphicx}		
\usepackage{dcolumn}		
\usepackage{bm}				
\usepackage[all]{hypcap}
\usepackage{xcolor}

\definecolor{grey}{rgb}{0.5,0.5,0.5}
\definecolor{lightgrey}{rgb}{0.9,0.9,0.9}

\usepackage{xspace}
\usepackage[version=3]{mhchem}
\newcommand{\deriv}{\operatorname{d}\!}
\newcommand{\e}{\mathrm{e}}
\newcommand{\ket}[1]{\ensuremath{{\left|#1\right\rangle}}\xspace}

\newcommand{\braOket}[3]{\ensuremath{\langle#1|#2|#3\rangle}\xspace}

\makeatletter
\renewcommand*{\p@section}{}
\renewcommand*{\p@subsection}{}
\renewcommand*{\p@subsubsection}{}
\makeatother

\begin{document}

\title{Comparison between 403~nm and 497~nm repumping schemes for strontium magneto-optical traps}

\author{P.H.~Moriya}	\altaffiliation{Current address: Institute of Photonics, Department of Physics, University of Strathclyde, The Technology and Innovation Centre, 99 George Street, Glasgow} \affiliation{Instituto de F\'isica de S\~ao Carlos, Universidade de S\~ao Paulo, 13560-970 S\~ao Carlos, SP, Brazil} 
\author{M.O.~Ara\'ujo}	\affiliation{Instituto de F\'isica de S\~ao Carlos, Universidade de S\~ao Paulo, 13560-970 S\~ao Carlos, SP, Brazil} \affiliation{Universit\'e C\^ote d'Azur, CNRS, Institut de Physique de Nice, France}
\author{F.~Tod\~ao}	\affiliation{Instituto de F\'isica de S\~ao Carlos, Universidade de S\~ao Paulo, 13560-970 S\~ao Carlos, SP, Brazil}
\author{M.~Hemmerling}	\affiliation{Instituto de F\'isica de S\~ao Carlos, Universidade de S\~ao Paulo, 13560-970 S\~ao Carlos, SP, Brazil}
\author{H.~Ke{\ss}ler}	\affiliation{Instituto de F\'isica de S\~ao Carlos, Universidade de S\~ao Paulo, 13560-970 S\~ao Carlos, SP, Brazil}
\author{R.F.~Shiozaki}	\affiliation{Departamento de F\'isica, Universidade Federal de S\~ao Carlos, Rodovia Washington Lu\'is, km 235 - SP-310, 13565-905 S\~ao Carlos, SP, Brazil}	\affiliation{Instituto de F\'isica de S\~ao Carlos, Universidade de S\~ao Paulo, 13560-970 S\~ao Carlos, SP, Brazil}
\author{R.~Celistrino Teixeira}	\affiliation{Departamento de F\'isica, Universidade Federal de S\~ao Carlos, Rodovia Washington Lu\'is, km 235 - SP-310, 13565-905 S\~ao Carlos, SP, Brazil}	\affiliation{Instituto de F\'isica de S\~ao Carlos, Universidade de S\~ao Paulo, 13560-970 S\~ao Carlos, SP, Brazil}
\author{Ph.W.~Courteille}\email[E-mail: ]{philippe.courteille@ifsc.usp.br}
\affiliation{Instituto de F\'isica de S\~ao Carlos, Universidade de S\~ao Paulo, 13560-970 S\~ao Carlos, SP, Brazil}

\begin{abstract}
The theoretical description of the external degrees of freedom of atoms trapped inside a magneto-optical trap (MOT) often relies on the decoupling of the evolution of the internal and external degrees of freedom. That is possible thanks to much shorter timescales typically associated with the first ones. The electronic structure of alkaline-earth atoms, on the other hand, presents ultra-narrow transitions and metastable states that makes such an approximation invalid in the general case. In this article, we report on a model based on open Bloch equations for the evolution of the number of atoms in a magneto-optical trap. With this model we investigate the loading of the strontium blue magneto-optical trap under different repumping schemes, either directly from a Zeeman slower, or from an atomic reservoir made of atoms in a metastable state trapped in the magnetic quadrupolar field. The fluorescence observed on the strong 461~nm transition is recorded and quantitatively compared with the results from our simulations. The comparison between experimental results and calculations within our model allowed to identify the existence of the decay paths between the upper level of the repumping transition and the dark strontium metastable states, which could not be explained by electric dipole transition rates calculated in the literature. Moreover, our analysis pinpoints the role of the atomic movement in limiting the efficiency of the atomic repumping of the Sr metastable states.
\end{abstract}

\maketitle

\section{Introduction}

The atomic dynamics inside a magneto-optical trap (MOT) is ruled by the evolution of the coupled internal and external degrees of freedom of each atom, subjected to the incoming laser light, the reemitted photons from the other atoms, a magnetic field gradient, and collisions. Due to the complex interplay between these effects, all models developed to predict the MOT behavior rely on simplifications of the exact equations of evolution for the observables, or have otherwise a limited range of validity in the parameters' space. Although a full quantum treatment of the internal and external degrees of freedom of the atoms in a simplified 1D model can predict the limiting temperature of a molasses \cite{Castin89}, it is far too costly in computational resources for modeling a full 3D MOT with finite-sized beams. Most of the theoretical and numerical models developed for describing MOTs assume that the timescales for the evolution of the internal degrees of freedom are much shorter than the timescales for the evolution of the external ones, and both are thus decoupled. The average light force exerted on the atoms is first calculated as a function of the steady-state solution of the Bloch equations, depending upon the local external parameters. The evolution of the global MOT parameters - size, temperature and atomic number - can then be calculated using a general statistical model for the atoms subjected to the average light force \cite{Townsend95, Lindquist92}. Alternatively, one can perform a numerical simulation of the evolution of the position of the atoms subjected to the average light force \cite{Wohlleben01}.

The decoupling of the internal and external degrees of freedom of the atoms breaks down when they present narrow and ultranarrow atomic transitions, as it is the case for the alkaline-earth atoms~\cite{Katori99}. These transitions imply a slow evolution of the internal degrees of freedom, and allow for the existence of metastable states, the lifetime of which can be bigger than the lifetime of the MOT itself~\cite{Yasuda04}. Moreover, when considering the effects of optical repumping from such metastable states, or from different hyperfine levels of the fundamental state, usually very small intensities are needed, due to the very long (or infinite) lifetime of the states. The dynamics of the atomic internal degrees of freedom under such small intensities is not necessarily faster than the evolution of the atomic position and momentum, and once more the decoupling of the timescales becomes a poor approximation.

The strong current interest in atomic strontium for metrological and interferometrical applications \cite{Ido03,Takamoto05,Derevianko11,Hinkley13,Bloom14,Poli11} motivates a quest for compact solutions for the laser sources needed to cool, trap and control strontium gases \cite{Poli06,Shimada13,Poli14,Pagett16}. Most experiments require, apart from a laser at 461~nm driving the strong blue cooling transition $(5\text{s}^2) ^1\text{S}_0\leftrightarrow (5\text{s}5\text{p}) ^1\text{P}_1$ (see Fig.~\ref{fig:0}), at least one "repumping" laser for recycling the population of atoms pumped into the metastable state $(5\text{s}5\text{p}) ^3\text{P}_2$, and sometimes a second repumper laser, which would recycle the atoms pumped into the metastable $(5\text{s}5\text{p}) ^3\text{P}_0$ (to which some atoms decay after being repumped from the $^3\text{P}_2$). Generally, the atoms are precooled in a MOT operated on the blue cooling transition, before being transferred to a "red MOT" operated on the narrow intercombination line the $^1S_0\leftrightarrow{^3P}_1$ at 689~nm, which allows to reach temperatures in the $\mu$K range. A usual choice for repumping atoms from the $^3P_2$ state is the transition $(5\text{s}5\text{p}) ^3\text{P}_2 \rightarrow (5\text{s}5\text{d}) ^3\text{D}_2$ at 497 nm. Unfortunately, laser diodes emitting in the green spectral range currently do not exist \cite{JiangLingrong16}. Light sources at 497 nm are realized by frequency-doubling infrared laser diodes, which is cumbersome and expensive. On the other hand, a previous study \cite{Stellmer09,Stellmer14} proposes a repumping transition near 403~nm that can be conveniently reached by Blu-ray laser diodes. The relative efficiency of both repumping schemes depends, as discussed on the leak to the $(5\text{s}5\text{p}) ^3\text{P}_0$ level. A popular choice to circumvent this limitation is to use light in resonance with the $(5\text{s}5\text{p}) ^3\text{P}_0 \rightarrow (5\text{s}6\text{s}) ^3\text{S}_1$ transition, at $707~$nm.

This article proposes a different theoretical model for describing the MOT dynamics, when the evolution of the internal degrees of freedom of the atoms cannot be adiabatically eliminated. It is based on modified Bloch equations, which will be henceforth referred to as global open Bloch equations (GOBE). Differently from the usual populations and coherences of the optical Bloch equations, in this model the population of each atomic level represents the average number of atoms in the MOT that are found in that particular level; and the coherences between two atomic levels indicate the average coherence of one atom, multiplied by the instantaneous number of atoms inside the MOT. The external degrees of freedom of the atoms will give rise to an inhomogeneous broadening of the relevant atomic transitions, through several different mechanisms. First, the main thermal speed of atoms is responsible for a spread of the natural resonances due to Doppler shifts; second, the magnetic moment of atoms under the quadrupolar magnetic field of the MOT creates a spatially inhomogeneous differential Zeeman slower between the lower and upper level of the transition; and finally, collisions between atoms. These inhomogeneities appear in our model as additional timescales and decay rates; other timescales will arise from the loading and loss rate of atoms from the MOT. Our GOBE take into account, thus, the effects of the external degrees of freedom on the calculation of the evolution of the internal ones, inverting, in a way, the standard approach of using the internal evolution of the atoms as an input for a calculation of the position and velocity dynamics of the atoms inside the MOT. We use this model to compare the effectiveness of the two repumping schemes aforementioned, that use 497 and 403~nm light. This alternative approach allows us to identify the role of different internal and external timescales on the evolution of the $^{88}\text{Sr}$ atoms in a MOT for the two repumping schemes. The calculation showed in this paper proves that the model proposed can be extremely useful for predicting the behavior of magneto-optical traps of other alkaline-earth atoms, and also for atoms with even more complex internal structure, under different cooling and repumping schemes.

The paper is organized as follows: In the first section, we present our GOBE model. In the second section, we compare our model to observations made in various experimental sequences by adjusting its free parameters. This comparison allows us to pinpoint various interesting quantities, such as the MOT loading rate, trap loss rates of atoms shelved in metastable states, branching ratios for the decay of optically pumped excited states serving as intermediate levels for the green and the Blu-ray repumper, and the effective Rabi frequencies of the repumping transitions. In the last section, we summarize and discuss our main results.

\section{Model}
\label{Section:Model}

The Bloch equations for the populations and coherences of a single atom subjected to one or more monochromatic laser beams is a set of linear, first order differential equations that can be cast in the form

\begin{equation}\label{eq:bloch1}
	\dot{\vec\rho}^{\,(1)}(t) = M^{(1)}\,\vec\rho^{\,(1)} (t).
\end{equation}

The components of the vector $\vec\rho^{\,(1)} (t)$, which we call here density vector, consists of all the components of the density matrix, i.e.~the populations $\rho_n^{(1)}(t) \equiv\rho_{nn}^{(1)} (t)$ and the coherences $\rho_{mn}^{(1)}(t)$. The matrix $M^{(1)}$ contains the Rabi frequencies, and the decay and dephasing rates to which the atom is subjected. The populations and coherences of the density matrix must satisfy at all times $\sum_n \rho_n^{(1)}(t) = 1$ and $|\rho_{mn}^{(1)} (t)|^2 \le \rho_m^{(1)}(t) \rho_n^{(1)}(t)$. 

Now, suppose we have $N(t)$ atoms in a MOT at a certain time $t$. We can define a global density matrix as the sum of the density matrices of all atoms, and then we define our global density vector as 

\begin{equation}\label{eq:vectorDef}
\vec\rho (t) = \sum_{j = 1}^{N(t)} \vec\rho^{\,(j)}(t).
\end{equation}

This vector has components $\rho_n (t) = \sum_{j = 1}^{N(t)} \rho_n^{(j)}(t)$ and $\rho_{mn} (t) = \sum_{j = 1}^{N(t)} \rho_{mn}^{(j)}(t)$, which we will call the global populations and coherences, respectively, or just simply populations and coherences. The populations of this new density vector represent the average number of atoms in the MOT at the corresponding atomic level; and the coherences are interpreted as the average coherences of all atoms within the MOT. We will search for a first-order differential equation for the evolution of the global density vector, which in the most general form can be written as

\begin{equation}\label{eq:bloch}
	\dot{\vec\rho} (t) = M(t) \,\vec\rho (t) + \vec{b}(t).
\end{equation}

We note that this equation is somehow different from Eq.~\ref{eq:bloch1}. First, it presents a non-homogeneous source term, the vector $\vec{b}(t)$, which we call the loading rate vector. It represents the atomic loading of the MOT, experimentally obtained from a surrounding vapor, or a slowed atomic beam. Since the number of atoms in the MOT is a function of time, these equations are open; in particular, the sum of all populations is not conserved anymore, since from Eq.~\ref{eq:vectorDef} it is clear that we have now $\sum_n \rho_n (t) = N(t)$. The general relation $|\rho_{mn} (t)|^2 \le \rho_m (t) \rho_n (t)$ is however still valid. Second, the matrix $M(t)$ now is allowed to depend on time. This generalization accounts for experiments in which the MOT parameters are dynamically changed, such as the intensity or detuning of the trapping and repumping beams. It is also useful if one wants to model the average movement of an atom within the MOT magnetic field gradient through the temporal modulation of the detuning of the different transitions, as will be discussed later. Finally, since the density vector $\vec\rho (t)$ represents the atoms inside the MOT region, the matrix $M(t)$ must also account for the loss rate of the atoms from the MOT, which can depend on the specific internal level the atom finds itself in. For an atomic ensemble with negligible loss rate, and no inhomogeneities in the atomic properties, one must have $M(t) = M^{(1)}$. However, for a complex system such as a MOT, with its magnetic field gradient, finite-size beams and collision between atoms, this, in general, will not be true. 

\bigskip

\begin{figure}[t]
	\centerline{\includegraphics[width=8.7 truecm]{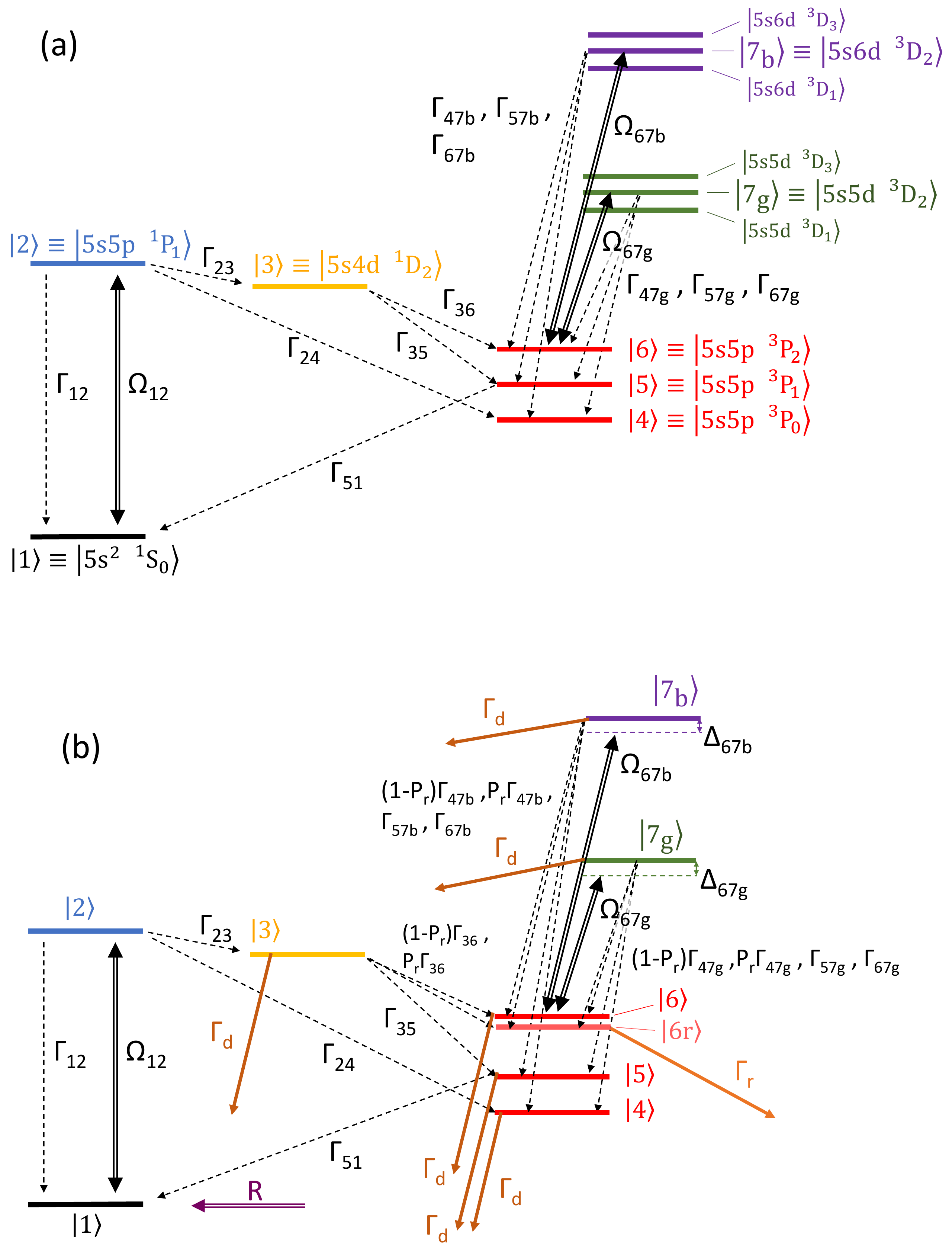}}
    \caption{(color online) (a): Strontium level diagram showing the transition frequencies and rates, as well as the Rabi frequency $\Omega_{12}$ due to the cooling MOT light, and $\Omega_{67g}$ and $\Omega_{67b}$ for the 497~nm and 403~nm repumper light, respectively. In this paper, simplified names are given for the relevant levels, such as $\ket{1}$ or $\ket{7_g}$, as explained in the main text. Note that the decay rate $\Gamma_{24}$ is not reported in the literature, but we have included following a hypothesis raised by Stellmer \textit{et al.}, \cite{Stellmer14}, discussed in the main text. See also table~\ref{table:1}. (b): All the relevant decay rates and timescales of the problem, as implemented in our model. The $\ket{6}$ level is split into two effective levels, $\ket{6}$ and $\ket{6r}$, with $\ket{6r}$ representing the atomic reservoir of atoms in the $(5\text{s}5\text{p}) ^3\text{P}_2$ electronic state that stays magnetically trapped in the quadrupolar magnetic field; $P_r$ is the probability to stay trapped. $\Gamma_d$ is the loss rate of atoms from the MOT region, that incurs on all atoms not trapped in the reservoir; and $\Gamma_r$ the much smaller decay rate of atoms from the reservoir. $\Delta_{67b,g}$ is the detuning of the repumping transition, which is a function of the atomic movement within the inhomogeneous magnetic field. The values used for the parameters are given in table~\ref{table:2}. Details in the main text.}
    \label{fig:0}
\end{figure}

We now apply these general concepts to the evolution of a $^{88}\text{Sr}$ MOT operating on the $\ket{1} \equiv (5\text{s}^2) ^1\text{S}_0\leftrightarrow \ket{2} \equiv (5\text{s}5\text{p}) ^1\text{P}_1$ transition. The relevant strontium levels and known decay rates are summarized in Fig.~\ref{fig:0} (a). In Fig. ~\ref{fig:0} (b), we show all levels that are implemented in our simulation, together with the effective parameters used, which we discuss in the following. An atomic beam decelerated by a Zeeman slower loads our MOT \cite{Moriya16}. We represent this loading by a vector $\vec{b}(t)$, whose components are all zero except for the one at the same position of the population $\rho_1$ of the ground state $\ket{1}$ at the $\vec \rho (t)$ vector, that is equal to the constant experimental loading rate $R$ when the Zeeman slower light beam is on; and $0$ otherwise. The excited state of the MOT transition may decay via the intermediate state $\ket{3} \equiv (5\text{s}4\text{d}) ^1\text{D}_2$ to either $\ket{5} \equiv (5\text{s}5\text{p}) ^3\text{P}_1$, which connects to the ground state via a weakly forbidden intercombination transition, or to the very metastable $\ket{6}~\equiv~^3\text{P}_2$ state. In order not to lose the atom in this state, it needs to be recycled by optical pumping via an excited state $\ket{7}$. Various repumping schemes have been used \cite{Bidel-01,Mickelson09,Kurosu92,Dinneen99,Poli05,Killian,Stuhler01}, the green transition at 497~nm addressing the $\ket{5} \rightarrow (5\text{s}5\text{d}) ^3\text{D}_2 \equiv \ket{7_g}$ transition being a popular choice \cite{Moriya16}. Stellmer \textit{et al.} \cite{Stellmer14} recently proposed to use the $\ket{5} \rightarrow (5\text{s}6\text{d}) ^3\text{D}_2 \equiv \ket{7_b}$ transition at 403~nm. The measurements described in the next section compare both schemes. It is important to note that both upper levels $\ket{7_g}$ and $\ket{7_b}$ present different decay probabilities to the metastable $(5\text{s}5\text{p}) ^3\text{P}_0 \equiv \ket{4}$ level by indirect dipole-allowed decay paths through several intermediate levels \cite{Stellmer14}. For $^{88}$Sr, the lifetime of the level is calculated to be thousands of years \cite{Santra04}; magnetic fields can shorten it to a fraction of a second \cite{Baillard07}. These atoms are thus lost from the trapping region. This limits the lifetime and maximum atomic number of the MOT.


All the relevant natural decay rates are given in table~\ref{table:1}. Most of them are direct natural decay rates, reported in the literature (the references are given in the table caption). For the $\Gamma_{k7g}$ and $\Gamma_{k7b}$ decay rates, with $k = 5,6$ (see fig. \ref{fig:0}), we have computed a total decay rate, considering the direct decay path and the indirect ones through dipole-allowed transitions, calculated following the lines of \cite{Stellmer14}. For the $\Gamma_{47g}$ and $\Gamma_{47b}$ decay rates, the above calculation gives $\Gamma_{47g} = 2\pi \times 3.3~$kHz and $\Gamma_{47b} = 2\pi \times 59~$kHz; however, these values are not corroborated by our experiments. Stellmer \textit{et al.} in ref. \cite{Stellmer14} had already reported divergences between the MOT behavior and the behavior that stems from the calculated decay rates. They hypothesized that a decay path from $\ket{2}$ directly to $\ket{4}$, still not reported in the literature, could be the explanation. We have thus included this possible additional decay path in our model, through the parameter $\Gamma_{24}$, supposed initially equal to zero. We will see that, in order to find agreement between the experimental results and the curves produced by our simulation, it is a modification of the $\Gamma_{47b}$ decay rate, instead of adding a $\Gamma_{24}$ decay rate, that is needed.

\begin{table}
\begin{tabular}[c]{|l|c|}
			\hline
			$\Gamma_{12}/2\pi$	& $30.5~\text{MHz}^{\text{a}}$ 	\\\hline
			$\Gamma_{23}/2\pi$	& $650~\text{Hz}^{\text{b}}$		\\\hline
			$\Gamma_{35}/2\pi$	& $352~\text{Hz}^{\text{c}}$		\\\hline
			$\Gamma_{36}/2\pi$	& $173~\text{Hz}^{\text{c}}$		\\\hline
			$\Gamma_{14}/2\pi$	& $\simeq 0^{\text{f}}$					\\\hline
			$\Gamma_{15}/2\pi$  & $7.5~\text{kHz}^{\text{d}}$		\\\hline
			$\Gamma_{16}/2\pi$	& $\simeq 0^{\text{e}}$					\\\hline
\end{tabular}
\quad
\begin{tabular}[c]{|l|c|}
			\hline
			$\Gamma_{24}/2\pi$		& $\simeq 0^{\text{h}}$					\\\hline
			$\Gamma_{47g}/2\pi$		& $3.3~\text{kHz}^{\text{g,i}}$	\\\hline
			$\Gamma_{57g}/2\pi$		& $7.15~\text{MHz}^{\text{g}}$	\\\hline
			$\Gamma_{67g}/2\pi$		& $2.25~\text{MHz}^{\text{g}}$	\\\hline
			$\Gamma_{47b}/2\pi$		& $59~\text{kHz}^{\text{g,i}}$	\\\hline
			$\Gamma_{57b}/2\pi$		& $3.26~\text{MHz}^{\text{g}}$	\\\hline
			$\Gamma_{67b}/2\pi$		& $1.04~\text{MHz}^{\text{g}}$	\\\hline
\end{tabular}
\caption{Table of decay rates found in the literature. These are the values used in our model, except for $\Gamma_{47g}$ and $\Gamma_{47b}$, that needed to be modified in order to match our experimental results, as discussed in the Measurements section. $^{\text{a}}$: \cite{Nagel05}. $^{\text{b}}$: \cite{Weiss86}, recalculated with the corrected value of $\Gamma_{12}$ from \cite{Nagel05}. $^{\text{c}}$: The branching ratio to the $\ket{5}$ and $\ket{6}$ levels are calculated in \cite{Bauschlicher85}; the total lifetime of the $\ket{3}$ level is measured in \cite{Vogel99}. $^{\text{d}}$: \cite{Drozdowski97}. $^{\text{e}}$: A calculated value of $\simeq 10^{-3}~$s$^{-1}$ is found in \cite{Santra04}; a measurement reported in \cite{Yasuda04} found $\simeq 2.10^{-3}~$s$^{-1}$ at $300~$K. These timescales are irrelevant, when compared to the background-pressure limited lifetime of our MOT. $^{\text{f}}$: A calculation reported in \cite{Santra04} suggests a natural lifetime of thousands of years. $^{\text{g}}$: Calculated oscillator strengths taken from \cite{Werij92}. From these, the total decay rate is calculated considering all possible dipole-allowed decay paths, following the lines of \cite{Stellmer14}, taking into account branching ratios and corrections from hyperfine splittings (as compiled in \cite{Sansonetti10}). $^{\text{h}}$: No decay path $\ket{2} \rightarrow \ket{4}$ is found in the literature, so we assume it to be zero. $^{\text{i}}$: Our experimental data do not agree with this value. See table~\ref{table:2} for the value that best fits our data.}
\label{table:1}
\end{table}

The cooling light of the MOT is red-detuned with respect to the atomic transition; moreover, due to the presence of a quadrupolar magnetic field, the transitions to different magnetic sublevels will have detunings that will be a function of the atomic position. Based on the spatial profile of the intensities of the cooling beams, the profile of the MOT atomic density and the magnetic field distribution, we calculate an average effective saturation parameter $s$ of the transition, and we use for our simulation a homogeneous effective Rabi frequency for the cooling light of $\Omega_{12} = \sqrt{s/2}\, \Gamma_{12}$~\cite{Barker15}.

The Rabi frequency of the 497~nm (403~nm) is $\Omega_{67g}$ ($\Omega_{67b}$). Due to the presence of the magnetic field gradient, the degeneracy among the magnetic sublevels of the lower and upper level of the repumping transition is lifted. This implies that several different transitions are now possible, each one with a different dipole moment, whose excitation depends on the polarization of the repumper light, and on the inhomogeneous quadrupolar magnetic field. We simplify the above picture, by neglecting the different magnetic sublevels and using an effective homogeneous Rabi frequency, calculated with the dipole moment of the hyperfine transition, $\braOket{6}{~\hat{\textbf{d}}~}{7_{b,g}}$. The inhomogeneous broadening, created by the differential Zeeman shift of the magnetic sublevels of the lower and upper states under the presence of a magnetic field gradient, is strong and cannot be neglected: For instance, the detuning of the transition $\ket{(5\text{s}5\text{p}) ^3\text{P}_2, m_J = 1} \rightarrow \ket{(5\text{s}5\text{d}) ^3\text{D}_2, m_J' = 2}$ has a sensibility to the magnetic field of $|g_J' m_J'-g_J m_J| \times (\mu_B/h) = 11/6 \times 2\pi \times 1.40~\text{MHz/G} = 2\pi \times 2.57~$MHz/G, with $\mu_B$ the Bohr magneton. For the gradient of magnetic field used for the production of the MOT, $70~\text{G/cm}$ in the coils' axis, we have a detuning gradient in this direction of = $2\pi \times 180~$MHz/cm.
Once the atoms decay out of the cooling cycle, to the $\ket{3}$ level and from there to the $\ket{5}$ or $\ket{6}$, they are free of light forces until they decay back to the $\ket{1}$ fundamental level. Due to their finite temperature, these free atoms will explore a volume, and thus a Zeeman broadening, bigger than the volume occupied by the atoms under the effect of the cooling beams.
To include this effect, we choose the detuning of the repumping transition $\Delta_{67b,g}$ with a probability distribution that averages the relative Zeeman shift of all possible magnetic sublevels and has a single free parameter: the spatial average excursion $\Delta r$ an atom makes from the center of the quadrupolar field before decaying back to the cooling cycle. Then, we average the results of our model with several $\Delta_{67b,g}$ chosen with the same probability distribution.

The absence of light forces when atoms decay out of the cooling cycle is also the main responsible for the atomic losses from the MOT in our low density regime. 
To account for this effect, a loss rate $\Gamma_d$ is added to all levels that are dark for the cooling light. However, some magnetic sublevels of the atomic states are trappable by the field configuration, and at longer timescales the atoms in these levels will still be around. This fact, combined with the long electronic stability of the metastable state $(5\text{s}5\text{p}) ^3\text{P}_2$, can actually be used to increase the number of atoms of the MOT, in a technique called reservoir loading \cite{Nagel03,Barker15}. The technique consists in accumulating the atoms that decay to the magnetically trappable sublevels of the $(5\text{s}5\text{p}) ^3\text{P}_2$ level in the magnetic trap formed by the quadrupolar magnetic field, during the blue MOT operation, instead of repumping them continuously. After a loading time of several seconds, which corresponds to the lifetime of the metastable atoms in the magnetic trap, a quick repumper pulse depletes the reservoir, sending them back to the fundamental state. The number of atoms obtained by this technique is often bigger than the number of atoms trapped in a continuously repumped MOT. To account for this effect, we have duplicated in our model the level $\ket{6} \equiv (5\text{s}5\text{p}) ^3\text{P}_2$ (see fig. \ref{fig:0}) into two levels, $\ket{6}$ and $\ket{6r}$. For any atomic level presenting a decay rate of $\Gamma_{6j}$ to the $(5\text{s}5\text{p}) ^3\text{P}_2$, we implement effective decay rates of $(1-P_r) \Gamma_{6j}$ and $P_r\,\Gamma_{6j}$ from this level to the $\ket{6}$ and $\ket{6r}$ levels, respectively. $P_r$ is the probability for an atom to stay trapped by the quadrupolar field. Additionally, the $\ket{6r}$ level will not have the loss rate $\Gamma_d$, but instead the lower decay rate $\Gamma_r$ that accounts for all loss mechanisms from which a magnetically trapped atom suffer (collisions with background gas, Majorana losses~\cite{Sukumar97}, collisions with other cold atoms within the trap).



The matrix $M(t)$ of eq. \ref{eq:bloch} will be a function of all the parameters described, and it is given explicitly in the Appendix. By assigning values to each one of them, we can integrate eq. \ref{eq:bloch} numerically and calculate the temporal evolution of all atomic populations. In our experiment, we measure the atomic fluorescence $F$ at the 461~nm atomic transition. For the low optical density of our MOT and the small saturation parameter of the trapping light, we can assume that this fluorescence is proportional to the number of atoms in the $\ket{2}$ state, 

\begin{equation}\label{eq:fluorescence}
F \propto \rho_2~.
\end{equation}

We have performed several different experimental sequences, in order to separate the effect of the different parameters discussed and determine the value of each one of them that best describes all our experimental data. This procedure is discussed in the next section, and the final values for the parameters are given in table~\ref{table:2}.

\section{Measurements}

In our experiment, we produce a collimated $^{88}$Sr atomic beam from a heated oven ejecting atoms through a two-dimensional array of microtubes. A 35~cm long spin-flip Zeeman slower (ZS) decelerates and injects the atoms into a standard MOT operated on the 461~nm cooling transition. The Zeeman slower laser beam has a power of $60~$mW, a waist of 5~mm at the entrance window and a waist of $2~$mm at the oven nozzle. It is tuned $-450~$MHz below the cooling transition. The cooling light of the MOT has a detuning of $\Delta = - 30~$MHz, and the magnetic field gradient created by a pair of anti-Helmholtz coils is equal to $70~$G/cm at the coils' axis. Each one of the six MOT beams has a waist of $w_{461} = 5~$mm, and the total intensity of the six beams at the center is equal to $2.1 I_{\text{sat}}$, where $I_{\text{sat}} = 40.5~$mW/cm$^2$. The atomic cloud at the steady state, with or without repumping light, is well approximated by a Gaussian symmetrical profile with a radius at $e^{-1/2}$ of $0.9~$mm. With these parameters, we calculate the average non-resonant saturation parameter seen by the atoms to be $s = 0.17$. As discussed previously, we then implement in our simulation an effective Rabi frequency $\Omega_{12}$ equal to $\sqrt{s/2}\, \Gamma_{12}$, and an effective detuning of zero, guaranteeing the same average atomic population in the excited level $\ket{2}$ in the experiment and the simulations. Two different repumping beams can be used, with wavelengths 497 and 403~nm, of waists $3.8$ and $3.6~$mm respectively. The MOT reaches a maximum number of atoms of $5.10^7$ atoms and has a temperature of $12~$mK.

We observe the blue MOT fluorescence in different experimental sequences switching on and off the blue MOT laser at 461~nm (MOT), the Zeeman slower laser at 461~nm (ZS), and either the green or the Blu-ray repumper laser (REP). The measurement of the atomic number through the MOT fluorescence is calibrated by an independent measurement of the atomic number by absorption imaging. We have performed three different experimental sequences: (A) MOT loading, (B) reservoir repumping, and (C) reservoir recapture efficiency, which are discussed in the following subsections.

\subsection{MOT loading}

\begin{figure}
	\centerline{\includegraphics[width=8.7 truecm]{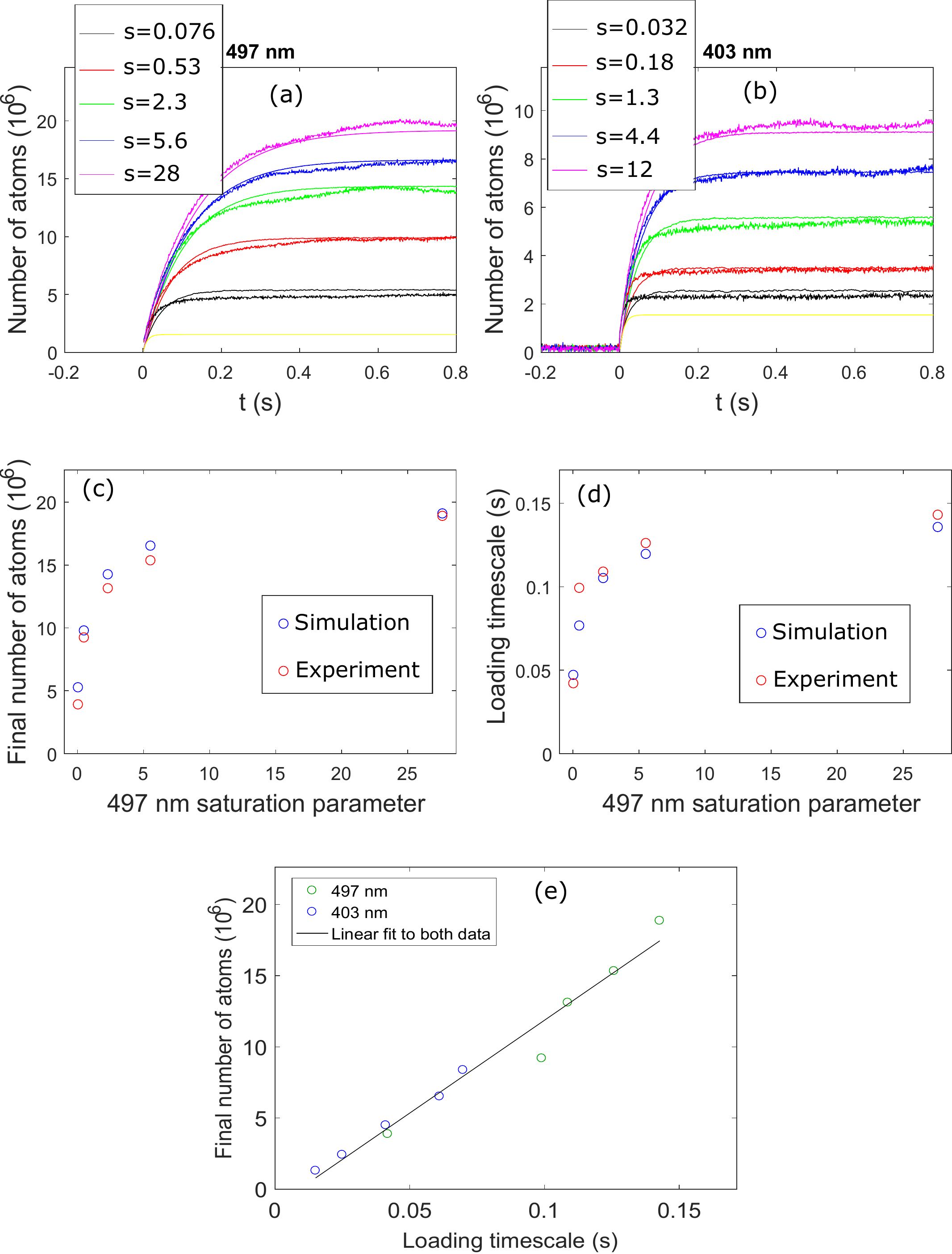}}
    \caption{(color online) Simulation of the MOT loading. (a) Total atom number in the MOT for different saturation parameters of the 497~nm repumper and (b) of the 403~nm repumper. The thick lines are experimental results, while the thin lines are the results of the simulation. (c) Final number of atoms and (d) loading timescale for the 497~nm repumper, as a function of the saturation parameter, for the experiments and simulation. (e) Final number of atoms as a function of the timescale, showing the linear relation between both parameters. The slope of this graphic is equal to the loading rate of our MOT, $R = 1.3 \times 10^8~$atoms/s.}
    \label{fig:A1}
\end{figure}

In a first series of measurements, we study the MOT loading behavior and the steady-state MOT fluorescence, as a function of the intensity of the repumping beams. The experimental sequence consists in operating the repumper (REP) and the Zeeman slower (ZS) continuously and recording the fluorescence after the MOT beams have been switched on. Figs.~\ref{fig:A1} (a) and (b) show typical loading curves measured when using 497 and 403~nm repumper beams, respectively, with a different intensity for each curve. The oscillations observed in some signals are due to technical imperfections, such as laser frequency instabilities.

The evolution of the atomic number is well described by an exponential behavior for all curves. If one can neglect many-body losses, the number of atoms of the MOT follow a simple temporal behavior of the form
\begin{equation}
\frac{\deriv N(t)}{\deriv t} = R - \frac{N(t)}{\tau}
\end{equation}
with $\tau$ the timescale for an atom to be lost from the trap. For a MOT that begins in $t = 0$ with $N = 0$, the solution of this equation is $N(t) = R \tau (1-\e^{-t/\tau})$ for $t >0$, with final atomic number $N_{\infty} \equiv N (t = \infty) = R \tau$. The main loss mechanism is assumed to be caused by atoms decaying to the metastable state $\ket{6}$. Our experimental results show that the timescale $\tau$ of the MOT loading is a function of the repumper intensity. In the absence of repumper light, we can make a rough estimate of the timescale $\tau_0$ as being equal to the average time the atoms spend in the cooling cycle times the probability that they are recycled back to the cooling cycle after decaying to the $\ket{3}$ level. The decay rate from the cooling cycle is estimated as the mean probability to be found at the excited level $\ket{2}$, $s/2(s+1)$, times the decay rate to the level $\ket{3}$, $\Gamma_{23}$; The probability to not return quickly to the cooling cycle after decaying to the $\ket{3}$ level is equal to $\Gamma_{36}/(\Gamma_{35} + \Gamma_{36})$, and thus

\begin{equation}
\tau_0 \simeq \left(\frac{1}{2}\frac{s}{s+1} \Gamma_{23}\right)^{-1} \frac{\Gamma_{36}}{\Gamma_{35} + \Gamma_{36}}.
\end{equation}

For the average saturation parameter seen by the atoms, we calculate $\tau_0 \simeq 10~$ms. Any increase in this timescale is due to a recycling effect of the repumping light, and together with this increase in timescale, we have an increase in the final number of atoms for the MOT through $N_\infty = R \tau$. For all curves in figs.~\ref{fig:A1} (a) and (b), we extract the final number of atoms $N_\infty$ and we show it in the graphic of fig. ~\ref{fig:A1} (e) as a function of the fitted exponential timescale $\tau$. The linear behavior shows that our MOT number is indeed limited by one-body losses. The slope gives the loading rate $R = 1.3 \times 10^8~$s$^{-1}$ of our MOT.

In our model, the behavior of the MOT final atomic number (or the MOT loading timescale) as a function of the saturation parameter of the 403 and 497~nm repumpers depends on the different decay rates of table~\ref{table:1}, on the global loss rate $\Gamma_d$ and on the inhomogeneous broadening of the repumping transition, which in our model depends only on the spatial average excursion $\Delta r$. Our first attempts on fitting our data assumed all calculated natural decay rates to be true, and thus considered as free parameters only $\Gamma_d$ and $\Delta r$. We have found that, with a given set of parameters, it is possible to fit the loading curves measured with either one of the to repumping schemes, but never for both at the same time. This is similar to what has been previously reported by Stellmer \textit{et al.} in ref.~\cite{Stellmer14}: there seems to be a discrepancy between the behavior predicted for the two repumper schemes when considering the measured and calculated loading rates of table~\ref{table:1}, and what is actually measured in an experiment.

To explain this discrepancy, Stellmer \textit{et al.} suggested the existance of a direct decay rate from the level $\ket{2}$ to the metastable level $\ket{4}$, from where they cannot be recycled with any of the two repumping schemes. We have tried this hypothesis, without success. The addition of a non-zero decay rate $\Gamma_{24}$ reduces the efficiency (i.e, by the final number of atoms of the MOT) of both 403 and 497~nm repumper schemes by the same factor for all loading curves. This is to be expected, since the proportion of atoms available at the level $\ket{6}$ is reduced by the same factor for both schemes in presence of a non-zero $\Gamma_{24}$. It is important to note that this conclusion by itself does not mean that our model can rule out the existence of a direct decay path from $\ket{2}$ to $\ket{4}$. Actually, the reduction of efficiency caused by $\Gamma_{24}$ can be perfectly compensated for by a modification in parameters $\Gamma_d$ and $\Delta r$. For instance, the loading curves for the 497~nm repumper are well reproduced with $\Gamma_{24} = 0$ by using $\Gamma_d = 100~\text{s}^{-1}$ and $\Delta r = 6~$mm; or instead, when supposing $\Gamma_{24} = 2\pi \times 10~Hz$, a similar result is found for $\Gamma_d = 40~\text{s}^{-1}$ and $\Delta r = 12~$mm. Since $1/\Gamma_d$ represents the timescale for atoms to escape from the region determined by the crossing of the cooling beams, a rough estimation of $\Gamma_d$ can be done by calculating the average time an atom takes to travel a distance equal to the cooling beams' waist. The quadratic thermal mean speed of atoms is equal to $v_{\text{rms}} = \sqrt{k_B T/m} = 1.1~$m/s for the temperature of $12~$mK of our MOT, with $k_B$ the Boltzmann constant and $m$ the mass of a $^{88}$Sr atom. An atom with this speed travels a distance of $w_{461}$ in $\Delta t = w_{461}/v_{\text{rms}} = 4.5~$ms, and we would expect that $\Gamma_d$ should be on the order of $1/\Delta t = 220~\text{s}^{-1}$. In the same way, an estimation of $\Delta r$ can just be qualitative, because it is an effective parameter that describes the average increase of the detuning and the reduction of the laser intensities as the atom explores a higher magnetic field away from the center; we can just say that it should be at most of the order of the repumper beams' waists. Thus, although the addition of $\Gamma_{24}$ is not enough to explain our experimental results, we cannot assign a precise value for it, apart from establishing a loose upper limit of tens of Hz, when $\Gamma_d$ and $\Delta r$ would become too unrealistic with respect to the physical processes that they presumably describe.

The relative behavior of both repumpers will only depend on the decay rates $\Gamma_{j7b,g}$, for $j = 4,5,6$; all other parameters affect both curves in the same proportion, and cannot explain the discrepancy. The decay rates shown in table~\ref{table:1} are calculated only considering dipole-allowed transitions between the upper triplet levels $\ket{7b,g}$ and the lower triplet levels \ket{4}, \ket{5} and \ket{6}, considering all possible decay paths as summarized in ref.~\cite{Stellmer14} (in particular, we note that although the electric dipole element between $\ket{7b,g}$ and $\ket{4}$ is equal to zero, $\Gamma_{47b,g}$ in table~\ref{table:1} are different from zero purely due to indirect dipole-allowed decay paths). The values of all transition strengths used in this calculation are found in ref.~\cite{Werij92}, obtained from numerical methods adapted to the calculation of valence spectra for alkaline-earth elements. A comparison between these calculated values and experimental measurements found in the literature grants a possible error of 10\% for them, which is not enough to explain our results. The off-resonance excitation rate of the $(5\text{s}n\text{d}) ^3\text{D}_1$ levels with repumper light is negligible: The detuning with respect to any transition to these levels is of at least $2 \pi \times 150~$GHz (which happens for the $(5\text{s}5\text{p}) ^3\text{P}_2 \rightarrow (5\text{s}6\text{d}) ^3\text{D}_1$ transition, excited by the 403~nm repumper) and the width of these levels is small, respectively $2 \pi \times 71~$kHz and $2 \pi \times 155~$kHz for $n = 5$ and $6$. With these parameters, our maximum laser intensities yield transition rates smaller than $10^{-4}~$s$^{-1}$. We thus attribute the discrepancy to spin-forbidden or dipole-forbidden transitions, not taken into account in the calculation of the transition rates of table~\ref{table:1}. Examples of spin-forbidden transitions are the $\ket{3} \rightarrow \ket{4}, \ket{5}$ and the $\ket{5} \rightarrow \ket{1}$ ones, that have non-zero matrix elements mainly due to a breakdown of the LS coupling \cite{Bauschlicher85}. This effect mixes the spin $S$ of the levels and allows for a small presence of a singlet character on triplet levels and \textit{vice versa}. It creates additional matrix elements between levels, that increase the decay rates by at most tens of kHz, which would only be appreciable for the small $\Gamma_{47b,g}$ decay rates. Since we are only sensitive to the relative behavior, we have kept in our simulations the value of $\Gamma_{47g}/2\pi = 3.3~\text{kHz}$ and modified $\Gamma_{47b}$ to $\Gamma_{47b}/2\pi = 190~\text{kHz}$. The theoretical curves of Figs.~\ref{fig:A1} (a) and (b) are calculated using these new parameters, together with the other values given in table~\ref{table:1}, $\Gamma_d = 100~\text{s}^{-1}$ and $\Delta r = 6~$mm. The agreement is now excellent. Figs.~\ref{fig:A1} (c) and (d) show the parameters $\tau$ and $N_{\infty}$ as a function of the saturation parameter for the 497~nm repumper, confirming the good agreement.

From the parameters found, we can also calculate the maximum increase of atoms that one can obtain with each one of the repumper schemes for our cooling beams' size: we predict an increase by a factor of $\simeq 13$ in the number of atoms with the 497~nm repumper, and $\simeq 8$ with the 403~nm repumper. We note that the available total power we had was not enough to see the saturation of the 403~nm repumper scheme. That would happen for a saturation parameter $s_b \gtrsim 20$, due to the inhomogeneities created by the magnetic field, while in our experiment we were limited to $s_b = 12$. The analysis carried out here also shows that one could still probably profit from an increase in the size of the cooling beams, while keeping the intensity constant, since this would reduce the parameter $\Gamma_d$ and consequently increase the maximum efficiency of the repumping schemes. An analysis of the MOT loading for different beams' size would point out the precise roles of $\Gamma_d$ and $\Gamma_{24}$, and possibly even assign a precise value for the last parameter.

\subsection{Repumper switch}

We now focus on the measurements concerning the possibility of magnetically trapping the atoms in the metastable state $\ket{6r}$. In a first series of experiments, we have measured the transient increase of the number of atoms of the MOT as a function of the loading time of the atomic reservoir (the atoms magnetically trapped in $\ket{6r}$). We first load our MOT with the cooling and repumping light on for $2~$s, and the Zeeman slower continuously on, which is enough to achieve a stationary atomic number. Then, we turn off the repumping light for a variable time $t_{\text{hold}}$. During this period, the fluorescence of the MOT quickly drops to a smaller stationary state, because the atoms that decay to the $\ket{6}$ level are not repumped back anymore. Some of the atoms in the $\ket{6}$ level stay trapped in the magnetic trap, in what we call the $\ket{6r}$ level. After the time $t_{\text{hold}}$, we turn on again the repumper light, and these atoms can be repumped back to the cooling cycle.

\begin{figure}[t]
	\centerline{\includegraphics[width=8.7 truecm]{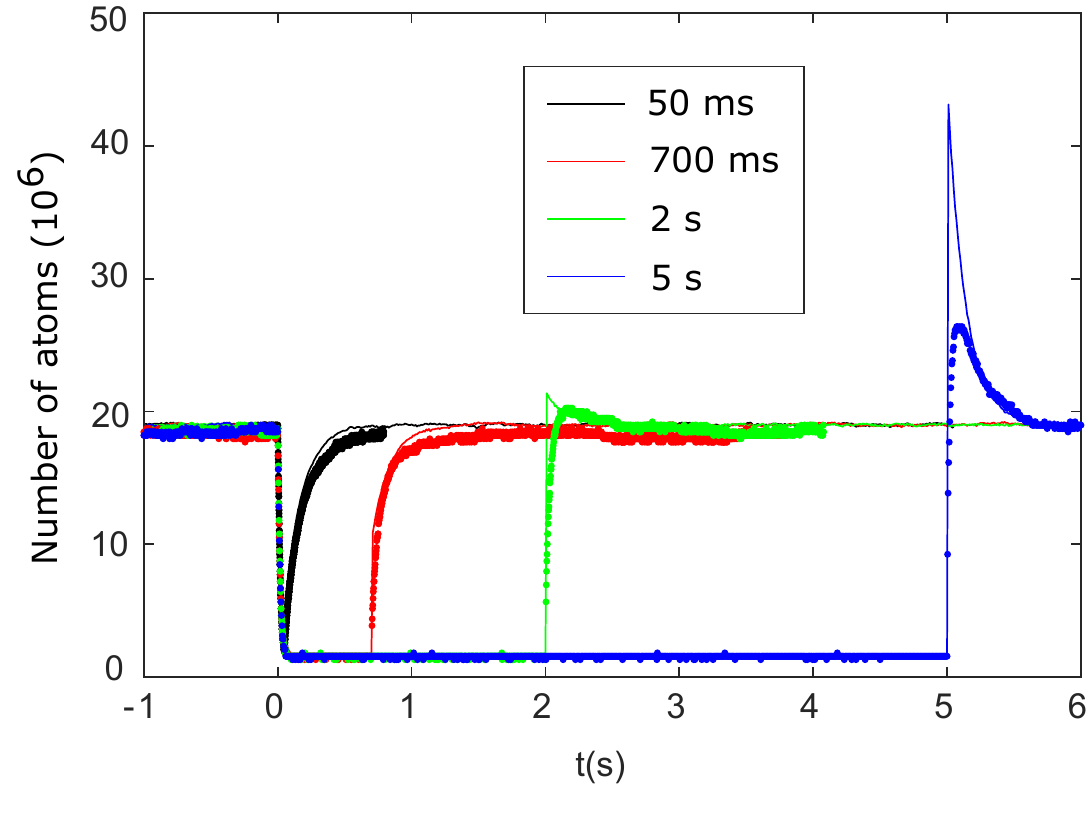}}
    \caption{(color online) Recovering of atoms from the magnetic trap of metastable $\ket{6r}$ atoms (the atomic reservoir) for different $t_{\text{hold}}$ (indicated in the figure legend). For this experiment, the Zeeman slower and the cooling light is always on; at $t = 0$, the 497~nm repumper is turned off, and after $t_{\text{hold}}$ it is turned on again. The thick lines are experimental, and the thin lines are the simulation results. Details and discussion in the main text.}
    \label{fig:C1}
\end{figure}

Figure \ref{fig:C1} shows our experimental data, superposed with the simulation results. At $t = 0$, the repumping light is turned off; after a variable $t_{\text{hold}}$, it is turned on again. We can see that for $t_{\text{hold}} \gtrsim 1~$s, the number of atoms of the MOT is bigger than the stationary number of atoms by a short transient time. This happens because we can trap more atoms in the reservoir than in the MOT; once they are repumped back, though, it decays to the stationary atomic number, for which the loading rate equals the loss rate. The decay timescale is equal to the loading timescale found before. 

Apart from the parameters already fixed from the measurement before, we have here two relevant parameters to use: $P_r$, the proportion of atoms decaying into the $\ket{6}$ level that stay trapped in the reservoir, and $\Gamma_r$, the lifetime of the reservoir. In fig.~\ref{fig:C1}, we show our experimental data, together with the results of our simulations, when using the 497~nm repumper with a power of $14.8~$mW (which corresponds to a saturation parameter of 28 at its center), all parameters as before, plus $P_r = 0.07$ and $\Gamma_r = 0.045~\text{s}^{-1}$. These parameters are simply calculated from the lifetime of the atoms in the reservoir (see also next subsection), and the minimum $t_{\text{hold}}$ that guarantees a transient overshoot on the recovered number of atoms from the reservoir, when compared to the steady state number of atoms of the MOT with repumper. We see, though, that the simulation clearly fails in predicting the peak of the transient of the number of atoms recovered. The transient depends on two different timescales: The timescale for repumping atoms out of the reservoir, and the timescale for the MOT to decay to its stationary number of atoms. The second one is the same $\tau$ found in the series of measurements for the MOT loading, and is equal to $140~$ms for the 497~nm repumper intensity used here; both experimental and simulated curves present it. The timescale for the repumping of the atoms from the reservoir, though, is clearly different in both cases. While the simulated curves show an extremely fast increase of the transient peak, in a time $< 1~$ms that we cannot resolve, our experiment shows $t_{\text{rep}} \simeq 320~$ms. This explains the discrepancy in the height of the transient peaks. We attribute the emergence of the timescale $t_{\text{rep}}$ found in the experiment to the fact that the atoms at the reservoir have time and energy to explore a bigger spatial volume, due to their temperature. The magnetic equipotential for the $m_j = 1$ sublevel of the $(5\text{s}5\text{p}) ^3\text{P}_2$ reservoir state that corresponds to an energy of $k_B T$ is an ellipsoid of radius $34~$mm to the directions of smaller magnetic field gradient, and $17~$mm at the axis of the magnetic field coils. So, the reservoir volume is much larger than the size of the trapping and repumping beams; also, the inhomogeneous Zeeman shift is extremely large at its borders. This effect is not described in our simulation, that supposes, as said, that they remain for short times at an average $\Delta_r = 6~$mm from the center, once decayed. This result is interesting, and explains why the efficiency of the reservoir loading is less important than one could expect, based solely on the number of atoms expected from the reservoir. We also note that the proportion $P_r = 0.07$ of trapped atoms found is smaller than what we could expect based on simple arguments: 2/5 of the magnetic sublevels of the $(5\text{s}5\text{p}) ^3\text{P}_2$ level are trappable. By the geometry of our coils and the science chamber, that determines the height of the magnetic potential, and the temperature of the cloud, we would expect that approximately half of the atoms would stay trapped after the evaporation of the most energetic atoms from the Maxwell-Boltzmann energy distribution with temperature $T$, which would give us $P_r \simeq 0.2$. This is linked to the fact that, for a longer timescale in repumping the atoms, the first appearance of the transient overshoot in our measurements happens for longer $t_{\text{hold}}$ when compared to the simulations, diminishing the effective value found for $P_r$ for our fitting procedure.

\subsection{Reservoir lifetime}

\begin{figure}[t]
	\centerline{\includegraphics[width=8.7 truecm]{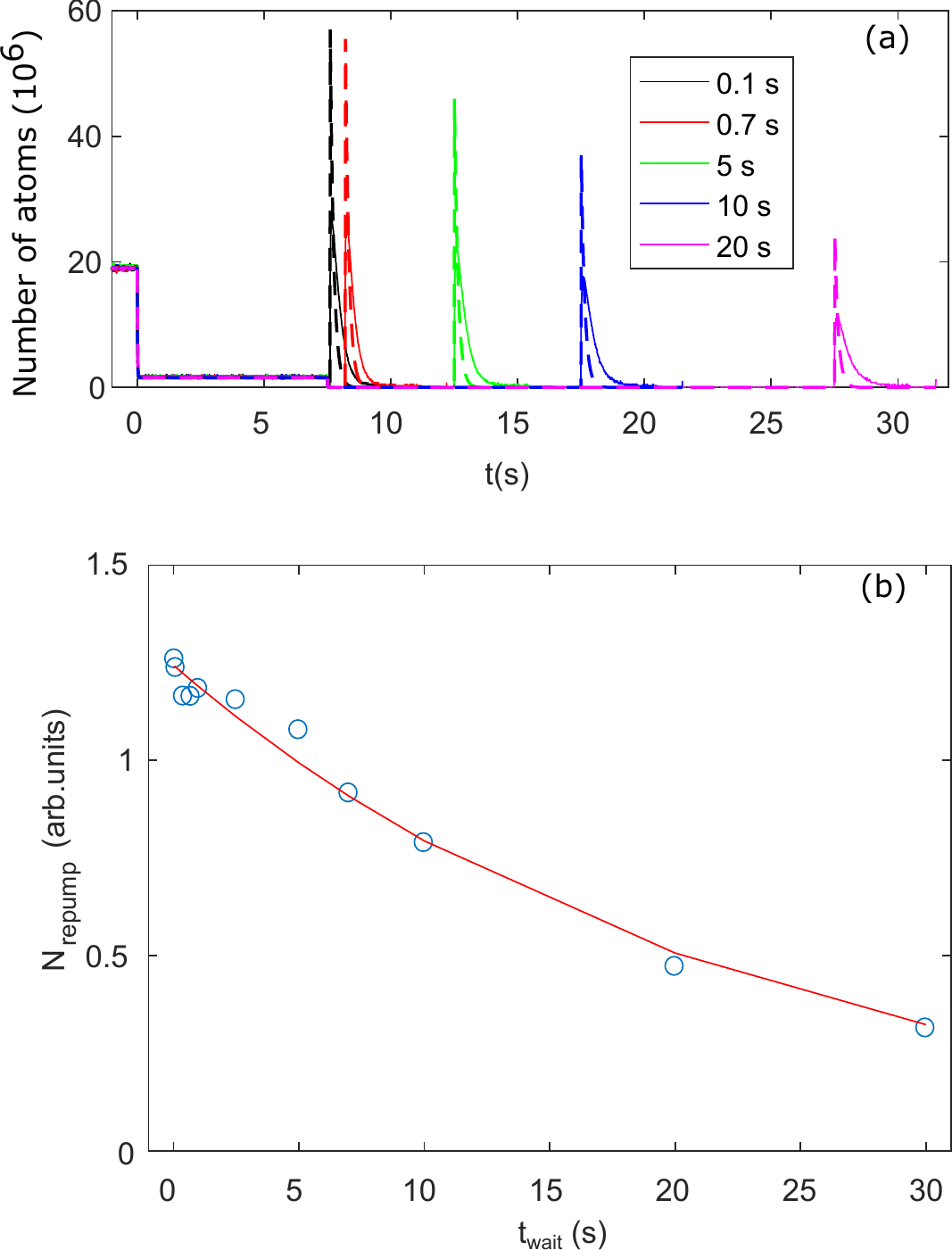}}
    \caption{(color online) (a) Measurement of the reservoir lifetime, with high visibility. The thin, continuous lines are experimental data, while the thick, dashed lines are the results of the simulations. For $t < 0$, the Zeeman slower, the cooling light of the MOT and the 497~nm repumper are turned on. At $t = 0$, the repumper is turned off; at $t = 7.5~s$, all lights are turned off, and after a variable time $t_{\text{wait}}$ indicated in the legend the cooling and repumper light are turned on again. (b) Maximum height $N_{\text{repump}}$ of the transient peak of fig. (a) as a function of the time $t_{\text{wait}}$. The solid red line is a fit of an exponential decay, giving a time constant of $t_{\text{res}} = 22,1~$s.}
    \label{fig:D}
\end{figure}

In a final series of measurements, we wanted to characterize the lifetime of the $\ket{6r}$ reservoir with maximum visibility. For that, we established the following procedure: We turn on the Zeeman slower, the cooling and the 497~nm repumping light (with a saturation parameter of $s = 28$ at the center of the cloud) until the MOT achieves steady state operation. Then, at $t = 0$ we turn off the 497~nm light, and let the atoms that decay from the cooling cycle accumulate in the $\ket{6r}$ reservoir for $7.5~$s. We then turn off all lights for a variable $t_{\text{wait}}$ time, and then abruptly turn on the cooling and 497~nm repumping light, in order to repump the atoms back to the cooling cycle, from there they eventually decay away after some time, completely depleting the MOT in absence of the atomic loading flux created by the Zeeman slower. Fig.~\ref{fig:D} (a) shows our experimental results, superposed to the results of our simulations. Here, the slower timescale $t_{rep} \simeq 320~$ms is found again on the slow decay of all experimental curves, limited by the repumping of atoms back to the cooling cycle. As discussed in the subsection before, this timescale cannot be recovered by our simulation; instead, the decay of the number of atoms in the simulation is governed again by the timescale for the losses of our MOT, $\tau = 140~$ms; this timescale is hidden in the longer timescale of $320~$ms in our experimental data. Although the transient peaks have different heights between experiment and simulation, the area of the total curve should be the same; this would give us a ratio between both peaks of $t_{\text{rep}}/\tau \simeq 2.3$, approximately found in the graphic of fig.~\ref{fig:D} (a).

The graphic of fig.~\ref{fig:D} (b) shows the peak height, $N_{\text{repump}}$, as a function of $t_{\text{wait}}$. From an exponential fit, we find the lifetime of the reservoir, $t_{\text{res}} = 22,1~$s, which allows us to calibrate the parameter $\Gamma_r$ already presented before. 

\begin{table}
\begin{tabular}[c]{|l|c|}
			\hline
			$\Omega_{12}/2\pi$		& $8,8~\text{MHz}$			\\\hline
			$\Gamma_{47b}/2\pi$		& $190~\text{kHz}$			\\\hline
			$\Gamma_d$						& $100~\text{s}^{-1}$		\\\hline
			$R$										& $1.3 \times 10^8~\text{s}^{-1}$			\\\hline
			
\end{tabular}
\quad
\begin{tabular}[c]{|l|c|}
			\hline
			$\Delta_r$						& $6~$mm								\\\hline
			$\Gamma_r$						& $0.045~\text{s}^{-1}$	\\\hline
			$P_r$									& $0.07$								\\\hline

\end{tabular}
\caption{Parameters used for our simulations, fitted from the experimental results. The decay rate $\Gamma_{47b}$ was modified from the value found at table~\ref{table:1}, in order to correctly explain our measurements, as discussed in the main text. The meaning of all parameters is discussed in section \ref{Section:Model}.}
\label{table:2}
\end{table}

\section{Discussion and conclusion}

We have set up a simple hybrid Bloch-rate equation model allowing us to simulate the evolution of the populations of the relevant states of atomic strontium trapped in a MOT. We have found that the main limitation of this model lies in the description of the atomic movement within the cloud. The atoms decay to metastable states, magnetically trappable or not, which induces spatial inhomogeneities that vary over time due to their movement and depends on their magnetic sublevel. We succeeded in implementing a simpler model, that reproduced quantitatively most of the features of the evolution curves for the atomic population of the MOT under different repumping intensities. One of the strongest conclusions of this work is the fact that the decay rate from the $(5\text{s}6\text{d}) ^3\text{D}_2$ to the level $(5\text{s}5\text{p}) ^3\text{P}_0$, compared to the decay rate from the $(5\text{s}5\text{d}) ^3\text{D}_2$ to the same $(5\text{s}5\text{p}) ^3\text{P}_0$ (or, in our notation, the ratio $\Gamma_{47b}/\Gamma_{47g}$) must be wrong by a factor of $\simeq 3$, when compared to the calculation considering only dipole-allowed transitions between triplet states. This has a direct impact on the evaluation of the efficiency of repumping schemes using both transitions, in the 403 and 497~nm wavelengths respectively, as it was shown here. The analysis presented here also allowed us to obtain the maximum theoretical repumping efficiency for our experimental parameters.

The study of the loading of the MOT from the atoms magnetically trapped in the metastable state $(5\text{s}5\text{p}) ^3\text{P}_2$ showed the limitations of the model, that fails in describing the effects of the slow motion of the atoms in the magnetic trap. But the comparison between the results of our model and the experimental data allowed us to pinpoint the reasons of the disagreement. It also allowed us to draw conclusions on the reasons for the smaller efficiency of this loading scheme, by a factor bigger than 2, when compared to a naive expectation considering only the proportion of atoms staying magnetically trapped and the lifetime of the experiment. 

The work described here can be seen as a proof of principle, and a calibration of our model. The parameters found here, and summarized in table~\ref{table:2}, can now be used to predict the behavior of our MOT under completely different conditions, which is of great help for the experimentalist. We note that the complexity of the level structure of alkaline-earth atoms, with the presence of singlet and triplet levels, metastable states and timescales different by several orders of magnitude, makes this model a useful tool for designing cooling schemes for these species.

Our future work will aim at studying the MOT operating at the intercombination narrow transition $(5\text{s}^2) ^1\text{S}_0 \leftrightarrow (5\text{s}5\text{p}) ^3\text{P}_1$, the so-called red MOT. A magneto-optical trap operating on a narrow transition is known to reach different temperature regimes, to have effects associated to the recoil momentum of the photon absorption, among other non-trivial effects~\cite{Katori99}. It will be intriguing to verify whether the model will be able to correctly predict the behavior of these MOTs under different experimental parameters.

\begin{acknowledgments}
This work was funded by grant 2013/04162-5, S\~ao Paulo Research Foundation (FAPESP), by the UK Engineering and Physical Sciences Research Council (EPSRC) under grant EP/I022791/1and the National Quantum Technology Hub for Sensors and Metrology (EP/M013294/1).
\end{acknowledgments}

\newpage
\begin{widetext}

\section{Appendix}


\newcommand{\G}[1]{\Gamma_{#1}}
\newcommand{\W}[1]{\frac{i}{2}\Omega_{#1}}
\newcommand{\K}[1]{-\Lambda_{#1}}
We define the Bloch vector by $\vec\rho\equiv\left(\rho_1~\rho_2~\rho_3~\rho_4~\rho_5~\rho_6~\rho_{6r}~\rho_7~\rho_{12}~\rho_{21}~\rho_{24}~\rho_{42}~\rho_{67}~\rho_{76}~\rho_{6r,7}~\rho_{7,6r}\right)^t$, in the notation of the main text. We note that most of the coherences $\rho_{jk}$ are always zero, because of absence of direct or indirect coherent coupling, and are not considered in the Bloch vector. The population dynamics can be described by the following matrix,
\begin{tiny}\begin{equation}
	M =  \left(
	\arraycolsep=0pt
	\begin{array}[c]{cccccccccccccc}
0		& \G{12}		& 0				& \G{14}		& \G{15}	& \G{16}		&\G{16} 	& 0				& \W{12} 	&-\W{12}	& 0			& 0		& 0		&0	\\
0		&-\G{12}-\G{23}-\G{24} & 0		&0			& 0				& 0			& 0				& 0				&-\W{12}	& \W{12}	& 0			& 0		& 0		& 0		\\
0		& \G{23}		&-\G{d}-\G{35}-\G{36}	& 0				& 0			& 0				& 0				& 0			& 0			& 0			& 0			&0		&0		&0\\
0		& \G{24}			& 0				&-\G{d}-\G{14}	& 0			& 0		&0		& \G{47}		& 0			& 0			& 0			& 0		&0		&0	\\
0		& 0				& \G{35}		& 0				&-\G{d}-\G{15}	& 0		&0		& \G{57}		& 0			& 0			& 0			& 0		&0		&0	\\
0 		& 0				& (1-P_r)\G{36}		& 0				& 0			&-\G{d}-\G{16} 	&0		& (1-P_r)\G{67}		& 0			& 0			& \W{67}	& -\W{67}		&0		&0\\
0 		& 0				& P_r\,\G{36}		& 0				& 0			&0		&-\G{r}-\G{16}		& P_r\,\G{67}		& 0			& 0			&0		&0		& \W{67}	& -\W{67}\\
0		& 0				& 0				& 0				& 0			& 0			&0			&-\G{d}-\G{47}-\G{57}-\G{67}& 0	& 0			& -\W{67}	& \W{67}	& -\W{67}	& \W{67}	\\
\W{12}	& -\W{12}		& 0		&0		& 0				& 0			& 0				& 0				& \K{12}	& 0			& 0			& 0		&0		&0	\\
-\W{12}	& \W{12}		& 0		&0		& 0				& 0			& 0				& 0				& 0			& \K{12}^*	& 0			& 0		&0		&0	\\
0		& 0				& 0				& 0				& 0			& \W{67}		&0		& -\W{67}		& 0			& 0			& \K{67} 	& 0		&0		&0\\
0		& 0				& 0				& 0				& 0			& -\W{67}		&0	& \W{67}		& 0			& 0			& 0			& \K{67}^*	&0		&0\\
0		& 0				& 0				& 0				& 0		 &0&		\W{67}		& -\W{67}		& 0			& 0			&0		&0		& \K{67} 	& 0\\
0		& 0				& 0				& 0				& 0		&0	& -\W{67}		& \W{67}		& 0			& 0			& 0		&0		&0		& \K{67}^*\\
\end{array}\right)
\end{equation}\end{tiny}
with $\Lambda_{jk}\equiv i\Delta_{jk}+\frac{1}{2}\Gamma_{jk}$, $\Delta_{jk}$ being the detuning between the quasi-resonant laser light and the $jk$ transition. We note that the matrix $M$ can depend upon the time through the Rabi frequencies $\Omega_{jk}$, that switch between a constant value and zero for describing the turning on and off of the different lasers, and through $\Delta_{67} (t)$,
that is calculated following a random trajectory of an atom inside a region of radius $\Delta_r$ around the center of the quadrupolar magnetic field, as described in the article.

\end{widetext}


\begin{thebibliography}{99}

\bibitem{Castin89}Y. Castin, H. Wallis, and J. Dalibard, JOSA B \textbf{11}, 2046 (1989).

\bibitem{Townsend95} C.G. Townsend, N.H. Edwards, C.J. Cooper, K.P. Zetie, C.J. Foot, A.M. Steane, P. Szriftgiser, H.Perrin, and J. Dalibard, Phys. Rev. A \textbf{52}, 1423 (1995).

\bibitem{Lindquist92} K. Lindquist, M. Stephens, and C. Wieman, Phys. Rev. A \textbf{46}, 4082 (1992).

\bibitem{Wohlleben01} W. Wohlleben, F. Chevy, K. Madison, and J. Dalibard, Eur. Phys. J. D \textbf{15}, 237 (2001).

\bibitem{Katori99}H. Katori, T. Ido, Y. Isoya, and M. Kuwata-Gonokami, Phys. Rev. Lett. \textbf{82}, 1116 (1999).

\bibitem{Yasuda04} M. Yasuda and H. Katori, Phys. Rev. Lett. \textbf{92}, 153004 (2004).

\bibitem{Ido03}T. Ido and H. Katori, Phys. Rev. Lett. \textbf{91}, 053001 (2003).

\bibitem{Takamoto05}M. Takamoto, F.-L. Hong, R. Higashi, and H. Katori, Nature (London) \textbf{435}, 321 (2005).

\bibitem{Derevianko11}A. Derevianko and H. Katori, Rev. Mod. Phys. \textbf{83}, 331 (2011).

\bibitem{Hinkley13}N. Hinkley, J.A. Sherman, N.B. Phillips, M. Schioppo, N.D. Lemke, K. Beloy, M. Pizzocaro, C.W. Oates, and A.D. Ludlow, Science \textbf{341}, 1215 (2013).

\bibitem{Bloom14}B.J. Bloom, T.L. Nicholson, J.R. Williams, S.L. Campbell, M. Bishof, X. Zhang, W. Zhang, S.L. Bromley, and J. Ye, Nature (London) \textbf{506}, 71 (2014).

\bibitem{Poli11}N. Poli, F.-Y. Wang, M.G. Tarallo, A. Alberti, M. Prevedelli, and G.M. Tino, Phys. Rev. Lett. \textbf{106}, 038501 (2011).


\bibitem{Poli06}N. Poli, G. Ferrari, M. Prevedelli, F. Sorrentino, R.E. Drullinger, G.M. Tino, Spectrochimica Acta Part A \textbf{63}, 981 (2006).

\bibitem{Shimada13}Y. Shimada, Y. Chida, N. Ohtsubo, T. Aoki, M. Takeuchi, Rev. of Sci. Instrum. \textbf{84}, 063101 (2013).

\bibitem{Poli14}N. Poli, M. Schioppo, S. Vogt, St. Falke, U. Sterr, Ch. Lisdat, and G.M. Tino, Appl. Phys. B \textbf{117}, 1107 (2014).

\bibitem{Pagett16}C.J.H. Pagett, P.H. Moriya, R.C. Teixeira, R. F. Shiozaki, M. Hemmerling, Ph.W. Courteille, Rev. Sci. Instr. \textbf{87}, 053105 (2016).

\bibitem{JiangLingrong16}Jiang Lingrong, et al., Journal Semicond. \textbf{37}, 111001 (2016), DOI:10.1088/1674-4926/37/11/111001.

\bibitem{Stellmer09}S. Stellmer, M.K. Tey, Bo Huang, R. Grimm, and F. Schreck, Phys. Rev. Lett. \textbf{103}, 200401 (2009).

\bibitem{Stellmer14}S. Stellmer and F. Schreck, Phys. Rev. A \textbf{90}, 022512 (2014). 

\bibitem{Moriya16}P.H. Moriya, R.F. Shiozaki, R.C. Teixeira, C.E. M\'aximo, N. Piovella, R. Bachelard, R. Kaiser, and Ph.W. Courteille, Phys. Rev. A \textbf{94}, 053806 (2016).

\bibitem{Mickelson09}P.G. Mickelson, Y.N. Martinez de Escobar, P. Anzel, B.J. DeSalvo, S.B. Nagel, A.J. Traverso, M. Yan, and T.C. Killian, J. Phys. B: At. Mol. Opt. Phys. \textbf{42}, 235001 (2009).

\bibitem{Kurosu92}T. Kurosu and F. Shimizu, Jpn. J. Appl. Phys. \textbf{31}, 908 (1992).

\bibitem{Dinneen99}T.P. Dinneen, K.R. Vogel, E. Arimondo, J.L. Hall, and A. Gallagher, Phys. Rev. A \textbf{59}, 1216 (1999).

\bibitem{Poli05}N. Poli, R.E. Drullinger, G. Ferrari, J. L\'eonard, F. Sorrentino, and G. M. Tino, Phys. Rev. A \textbf{71}, 061403 (2005).

\bibitem{Killian}T.C. Killian (private communication).

\bibitem{Stuhler01}J. Stuhler, P.O. Schmidt, S. Hensler, J. Werner, J. Mlynek, and T. Pfau, Phys. Rev. A \textbf{64}, 031405(R) (2001).

\bibitem{Bidel-01}Y. Bidel, PhD-thesis, Universit\'e de Nice - Sophia Antipolis (2001): \textit{Pi\'egeage et refroidissement laser du strontium: Etude de l'effet des interf\'erences en diffusion multiple}.

\bibitem{Santra04} R. Santra, K.V. Christ, and C.H. Greene, Phys. Rev. A \textbf{69}, 042510 (2004).

\bibitem{Baillard07} X. Baillard, M. Fouché, R. Le Targat, Ph. G. Westergaard, A. Lecallier, Y. Le Coq, G. D. Rovera, S. Bize, and P. Lemonde, Opt. Lett. \textbf{32}, 1812 (2007).

\bibitem {Barker15}, D.S. Barker, B.J. Reschovsky, N.C. Pisenti, and G.K. Campbell. Phys. Rev. A \textbf{92}, 043418 (2015).

\bibitem{Nagel05}S.B. Nagel, P.G. Mickelson, A.D. Saenz, Y.N. Martinez, Y.C. Chen, T.C. Killian, P. Pellegrini, and R. Côté, Phys. Rev. Lett. \textbf{94}, 083004 (2005).

\bibitem{Weiss86} L.R. Hunter, W.A. Walker, and D.S. Weiss, Phys. Rev. Lett. \textbf{56}, 823 (1986).

\bibitem{Bauschlicher85} C.W. Bauschlicher Jr., S.R. Langhoff, and H. Partridge, J. Phys. B: At. Mol. Phys. \textbf{18}, 1523 (1985).

\bibitem{Vogel99} K.R. Vogel, PhD thesis. University of Colorado, Boulder (1999). 

\bibitem{Drozdowski97} R. Drozdowski, M. Ignaciuk, J. Kwela, J. Heldt, Z. Phys. D \textbf{41}, 125 (1997).

\bibitem{Werij92} H.G.C. Werij, C.H. Greene, C.E. Theodosiou, and A. Gallagher, Phys. Rev. A \textbf{46}, 1248 (1992).

\bibitem{Sansonetti10} J.E. Sansonetti and G. Nave, J. Phys. Chem. Ref. Data \textbf{39}, 033103 (2010).

\bibitem{Nagel03} S.B. Nagel, C.E. Simien, S. Laha, P. Gupta, V.S. Ashoka, and T.C. Killian, Phys. Rev. A \textbf{67}, 011401(R) (2003).

\bibitem {Sukumar97} C.V. Sukumar, D.M. Brink, Phys. Rev. A \textbf{56}, 2451 (1997).










\end{thebibliography}

\end{document}